\documentclass[11pt]{article}
\usepackage[toc,page]{appendix}
\usepackage[utf8]{inputenc} 
\usepackage{geometry} 
\usepackage{graphicx} 
\usepackage{afterpage}
\usepackage{amsmath}
\usepackage{bm} 
\usepackage[strict]{changepage}
\usepackage[english]{babel}
\usepackage{natbib} 
\usepackage{amsfonts}
\usepackage{hyphenat}
\usepackage{booktabs} 
\usepackage{enumerate}
\usepackage{kbordermatrix} 
\usepackage{array} 
\usepackage{paralist} 
\usepackage[group-separator={,}]{siunitx}
\usepackage{verbatim}
\usepackage{subfig} 
\usepackage{setspace}
\usepackage{caption}
\usepackage{lscape}
\usepackage{tabu}
\usepackage[flushleft]{threeparttable}
\usepackage{fancyhdr} 
\usepackage{sectsty}
\usepackage[nottoc,notlof,notlot]{tocbibind} 
\usepackage[titles,subfigure]{tocloft} 

\title{\textbf{Efficient Maximum-Likelihood Inference For The Isolation-With-Initial-Migration Model With Potentially Asymmetric Gene Flow}}
\author{\small Rui J. Costa and Hilde Wilkinson-Herbots\\
\small Department of Statistical Science, \small University College London\\
\small Gower Street, London WC1E 6BT, UK\\
\small email: rui.costa.11@ucl.ac.uk
}
\date{} 

\begin{document}
\maketitle

\begin{abstract}
The isolation-with-migration (IM) model is a common tool to make inferences about the presence of gene flow during speciation, using polymorphism data.  However, \citet{Becquet2009} report that the parameter estimates obtained by fitting the IM model are very sensitive to the model's assumptions -- including the assumption of constant gene flow until the present. This paper is concerned with the isolation-with-initial-migration (IIM) model of \citet{Herbots2012}, which drops precisely this assumption. In the IIM model, one ancestral population divides into two descendant subpopulations, between which there is an initial period of gene flow and a subsequent period of isolation. We derive a fast method of fitting an extended version of the IIM model, which allows for asymmetric gene flow and unequal subpopulation sizes. This is a maximum-likelihood method, applicable to observations on the number of segregating sites between pairs of DNA sequences from a large number of independent loci.  In addition to obtaining parameter estimates, our method can also be used to distinguish between alternative models representing different evolutionary scenarios, by means of likelihood ratio tests. We illustrate the procedure on pairs of Drosophila sequences from approximately 30,000 loci. The computing time needed to fit the most complex version of the model to this data set is only a couple of minutes. The R code to fit the IIM model can be found in the supplementary files of this paper.
\end{abstract}

\begin{adjustwidth}{3.5em}{3.5em}
\end{adjustwidth}

\section{Introduction}

The 2-deme isolation-with-migration (IM) model is a population genetic model in which, at some point in the past, an ancestral population divided into two subpopulations. After the division, these subpopulations exchanged migrants at a constant rate until the present. The IM model has become one of the most popular probabilistic models in use to study genetic diversity under gene flow and population structure. Although applicable to populations within species, many researchers are using it to detect gene flow between diverging populations and to investigate the role of gene flow in the process of speciation. A meta-analysis of published research articles that used the IM model in the context of speciation can be found in  \citet{Pinho2010}.

Several authors have developed computational methods to fit IM models to real DNA data. Some of the most used programs are aimed at data sets consisting of a large number of sequences from a small number of loci. This is the case of \textit{MDIV} \citep{Nielsen2001}, \textit{IM} \citep{Hey2004,Hey2005}, \textit{IMa} \citep{Hey2007} and \textit{IMa2} \citep{Hey2010}, which rely on Bayesian MCMC methods to estimate the model parameters and are computationally very intensive.

In the past decade, the availability of whole genome sequences has increased significantly. Very large data sets consisting of thousands of loci, usually from just a few individuals, became available for analysis, and  such multilocus data sets are more informative than their single locus counterparts. In fact, as the sample size for a single locus increases, the probability that an extra sequence adds a deep (i.e. informative) branch to the coalescent tree quickly becomes negligible \citep[see, e.g.,][p. 28-29]{Hein2005}. This new type of data set is also much more suitable for likelihood inference: if at each locus the observation consists only of a pair or a triplet of sequences, the coalescent process of these sequences is relatively simple and can more easily be used to derive the likelihood for the locus concerned. Furthermore, if the loci studied are distant from each other, observations at different loci can be considered independent and thus the likelihood of a whole set will consist of the product of the likelihoods for the individual loci.

For these reasons, some of the most recent theoretical papers and computer implementations of the IM model are aimed at data sets consisting of a small number of sequences at a large number of loci. \citet{Wang2010} and \citet{Zhu2012}, for example, developed maximum-likelihood methods based on numerical integration. Computationally less intensive methods of fitting the IM model were studied by \citet{Lohse2011}, who used moment generating functions, and \citet{Andersen2014}, who resorted to matrix exponentiation and decomposition.

Despite all the progress, several authors have recently pointed out some limitations of the IM model. \citet{Becquet2009} report that the parameter estimates obtained by fitting the IM model are highly sensitive to the model's assumptions -- including the assumption of constant gene flow until the present.  \citet{Strasburg2011} and \citet{Sousa2011} note that the gene flow timing estimates reported in the literature, based on the IM model and obtained with the \textit{IMA2} program, have extremely wide confidence intervals, as the time of gene flow is non-identifiable in the IM model. 

As a step to overcome these limitations, \citet{Herbots2012} studied an extension of the IM model, the isolation-with-initial-migration (IIM) model, which is more realistic than the IM model in the context of speciation. Broadly speaking, the IIM model is an IM model in which gene flow ceased at some point in the past. Explicit formulae for the distribution of the coalescence time of a pair of sequences, and the distribution of the number of nucleotide differences between them, are derived in \citet{Herbots2012}. These analytic results enable a very fast computation of the likelihood of a data set consisting of observations on pairs of sequences at a large number of independent loci. However, for mathematical reasons this work was limited to the case of symmetric migration and equal subpopulation sizes during the migration period. 

In this paper, we study a more general IIM model which allows for asymmetric gene flow during the migration period. It also allows for unequal subpopulation sizes during gene flow, as well as during the isolation stage. Both this model and other simpler models studied in this paper assume haploid DNA sequences, which accumulate mutations according to the infinite sites assumption \citep{Watterson1975}. An extension to the Jukes-Cantor model of mutation is feasible but beyond the scope of this paper.

We first describe, for different versions of the IIM model, an efficient method to compute the likelihood of a set of observations on the number of different nucleotides between pairs of sequences (also termed the number of pairwise differences). Each pair of sequences comes from a different locus and we assume free recombination between  loci and no recombination within loci. Secondly, we illustrate how to use this method to fit the IIM model to real data. The data set of Drosophila sequences from \citet{Wang2010}, containing over 30,000 observations (i.e. loci), is used for this purpose. Finally, we demonstrate, using this data set, how different models, representing different evolutionary scenarios, can be compared using likelihood ratio tests.  

\section{Theory and methods}

\begin{figure} 
\graphicspath{ {./} }
\centering         
\includegraphics[width=8.5cm, angle=0]{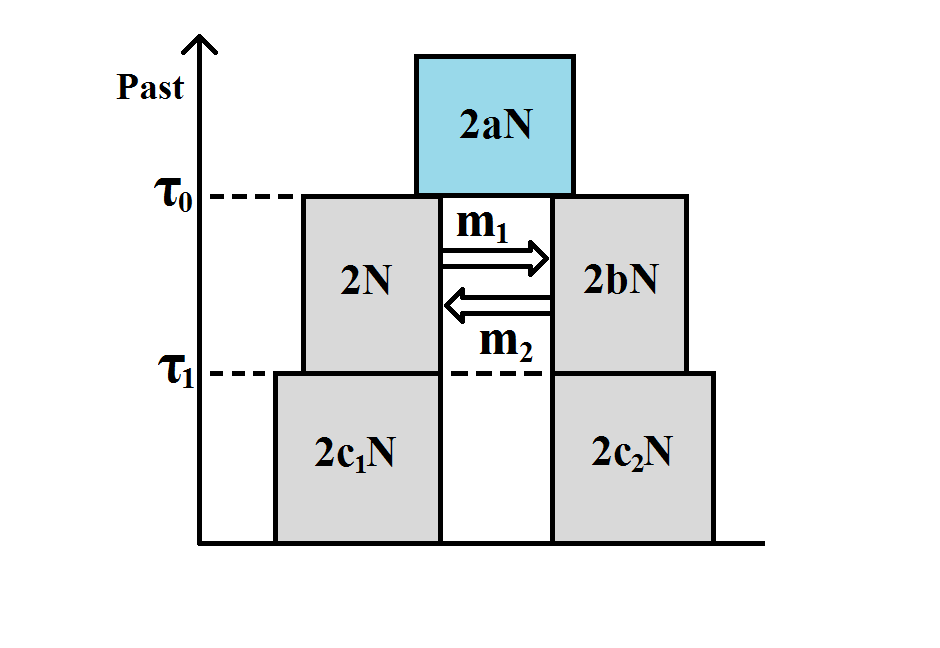} 
\caption{The IIM model. The left-hand side subpopulation is subpopulation 1; the right-hand side subpopulation is subpopulation 2.}      
\label{fig:IIMfull1} 
\end{figure}

For the purposes of the present paper, and from a forward-in-time perspective, the isolation-with-migration  (IM) model makes the following assumptions: a) until time $\tau_{0}$ ago ($ \tau_{0}>0 $), a population of DNA sequences from a single locus followed a Wright-Fisher haploid model \citep{Fisher1930,Wright1931}; b) at time $\tau_{0}$ ago, this ancestral population split into two Wright-Fisher subpopulations with constant gene flow between them. If we take an IM model and add the assumption that, at time $\tau_{1}$ ago ($0<\tau_{1}< \tau_{0}$), gene flow ceased, we get an isolation-with-initial-migration (IIM) model. Figure \ref{fig:IIMfull1} illustrates the fullest IIM model dealt with in this paper. 

In the IIM model of Figure \ref{fig:IIMfull1}, the population sizes are given inside the boxes, in units of DNA sequences. All population sizes are assumed constant and strictly positive. The parameters $a$, $b$, $c_{1}$ and $c_{2}$ indicate the relative size of each population with respect to subpopulation 1 during the migration stage. For example, if $2N_{anc}$ is the number of sequences in the ancestral population, then $a=2N_{anc}/2N$. Between times $\tau_{0}$ and $\tau_{1}$ ago (two time parameters in units of $2N$ generations) there is gene flow between the subpopulations: in each generation, a fraction $m_{i}$ of subpopulation $i$ are immigrants from subpopulation $j$ ($i,j \in \left\lbrace 1,2\right\rbrace$ with $i \neq j$), i.e. $m_{i}$ is the migration rate  per generation from subpopulation $i$ to subpopulation $j$ backward in time. Within each subpopulation, reproduction follows the neutral Wright-Fisher model and, in each generation, restores the subpopulations to their original sizes, i.e. reproduction undoes any decrease or increase in size caused by gene flow (this assumption of constant population size is common and usually reflects restrictions on food and habitat size).

Under the IIM model, the genealogy of a sample of two DNA sequences from the present subpopulations can be described by successive Markov chains, working backward in time. We will define these in the simplest possible way, using the smallest state space necessary for the derivation of the coalescence time distribution. Hence, during the isolation stage (until time $\tau_{1}$ into the past) and the migration stage (between $\tau_{1}$ and $\tau_{0}$), the process can only be in state $1$ -- both lineages in subpopulation 1 --, state $2$ -- both lineages in subpopulation 2 --, state $3$ -- one lineage in each subpopulation --, or state $4$ -- in which lineages have coalesced. After $\tau_{0}$, the lineages have either coalesced already -- state 4 --, or have not -- state 0. Only states 1, 2 and 3 can be initial states, according to whether we sample two sequences from subpopulation 1, two sequences from subpopulation 2, or one sequence from each subpopulation. When the genealogical process starts in state $i$ (with $i\in \left\lbrace \text{1,2,3} \right\rbrace$), the time until the most recent common ancestor of the two sampled sequences is denoted $T^{(i)}$, whereas $S^{(i)}$ denotes the number of nucleotide differences between them.  

If time is measured in units of $2N$ generations and $N$ is large, the genealogical process is well approximated by a succession of three continuous-time Markov chains, one for each stage of the IIM model \citep{Kingman1982coal,Kingman1982,Notohara1990}. We refer to this stochastic process in continuous time as the \textit{coalescent} under the IIM model. During the isolation stage, the approximation is by a Markov chain defined by the generator matrix 
\begin{equation}
\label{matrix:Qiso}
\begin{array}{lcl}
\renewcommand{\arraystretch}{1.5}
\mathbf{Q^{\left(\mathit{i}\right)}_{iso}}&=&\kbordermatrix{~&(i)&(4)\\
(i)&-\frac{1}{c_{i}}&\frac{1}{c_{i}}\\
(4)&0&0},\\
\end{array}
\end{equation}
with $i \in \left\lbrace 1,2\right\rbrace$ being the initial state \citep{Kingman1982coal,Kingman1982}. If 3 is the initial state, the lineages cannot coalesce before $\tau_{1}$. During the ancestral stage, the genealogical process is approximated by a Markov chain with generator matrix
 
\begin{equation}
\label{matrix:Q3}
\renewcommand{\arraystretch}{1.5}
\mathbf{Q_{anc}}=\kbordermatrix{~&(0)&&(4)\\
				(0)&-\frac{1}{a}&&\frac{1}{a}\\
				(4)&0&&0}.
\end{equation}
\citep{Kingman1982coal,Kingman1982}. In between, during the migration stage, the approximation is by a Markov chain with generator matrix 
\small
\begin{equation}
\label{matrix:Q2}
\renewcommand{\arraystretch}{1.5}
\text{\normalsize $\mathbf{Q_{mig}}$}=
\begin{array}{c}
\kbordermatrix{~&(1)&(3)&(2)&(4)\\
				(1)&-(1+M_{1})&M_{1}&0&1\\
				(3)&\frac{M_{2}}{2}&-\left(\frac{M_{1}+M_{2}}{2}\right)&\frac{M_{1}}{2}&0\\
			    (2)&0&M_{2}&-(1/b+M_{2})&1/b\\
			    (4)&0&0&0&0}
\end{array}
\end{equation}
\normalsize
\citep{Notohara1990}. In this matrix, $M_{i}/2=2Nm_{i}$ represents the rate of migration (in continuous time) of a single sequence when in subpopulation $i$. The rates of coalescence for two lineages in subpopulation 1 or 2 are $1$ and $1/b$ respectively. Note that state 3 corresponds to the second row and column, and state 2 to the third row and column. This swap was dictated by mathematical convenience: the matrix  $\mathbf{Q_{mig}}$ should be as symmetric as possible because this facilitates a proof in the next section.

\vspace{0.5cm}

\subsection{Distribution of the time until coalescence under bidirectional gene flow ($M_{1}>0$, $M_{2}>0$)}
\label{sec:dbn_coal_time_bi_mig}
To find $f^{\left(i\right)}_{T}$, the density of the coalescence time $T^{\left(i\right)}$ of two lineages under the IIM model, given that the process starts in state $i$ and there is gene flow in both directions, we consider separately the three Markov chains mentioned above. We let $T^{(i)}_{iso}$ ($i \in \{1,2\}$), $T^{(i)}_{mig}$ ($i \in \{1,2,3\}$) and $T^{(0)}_{anc}$ denote the times until absorption of the time-homogeneous Markov chains defined by the generator matrices $\mathbf{Q^{(i)}_{iso}}$, $\mathbf{Q_{mig}}$ and $\mathbf{Q_{anc}}$ respectively. And we let the corresponding \textit{pdf}'s (or \textit{cdf}'s) be denoted by $f^{(i)}_{iso}$, $f^{(i)}_{mig}$ and $f^{(0)}_{anc}$ (or $F^{(i)}_{iso}$, $F^{(i)}_{mig}$ and $F^{(0)}_{anc}$). Then $f^{\left(i\right)}_{T}$ can be expressed in terms of the distribution functions just mentioned:
\begin{equation}
\label{eq:general coal dens}
f^{\left(i\right)}_{T}(t) = \left\{
  \begin{array}{l l}
   f^{\left(i\right)}_{iso}(t)& \quad \text{for $0 \leq t \leq \tau_{1}$},\\
   \left[1-F^{\left(i\right)}_{iso}(\tau_{1})\right]f^{\left(i\right)}_{mig}(t-\tau_{1})  & \quad \text{for $\tau_{1} < t \leq \tau_{0}$},\\
   \left[1-F^{\left(i\right)}_{iso}(\tau_{1})\right]\left[1-F^{\left(i\right)}_{mig}(\tau_{0}-\tau_{1})\right]f^{\left(0\right)}_{anc}(t-\tau_{0}) & \quad \text{for $\tau_{0} < t< \infty$},\\
0 &\quad \text{otherwise}.
  \end{array} \right.
\end{equation}
for $i \in \left\lbrace 1,2 \right\rbrace$. If 3 is the initial state,
\begin{equation}
\label{eq:general_coal_dens_state3}
f^{\left(3\right)}_{T}(t) = \left\{
  \begin{array}{l l}
   f^{\left(3\right)}_{mig}(t-\tau_{1})  & \quad \text{for $\tau_{1} < t \leq \tau_{0}$},\\
   \left[1-F^{\left(3\right)}_{mig}(\tau_{0}-\tau_{1})\right]f^{\left(0\right)}_{anc}(t-\tau_{0}) & \quad \text{for $\tau_{0} < t< \infty$},\\
0 &\quad \text{otherwise}.
  \end{array} \right.
\end{equation}
The important conclusion to draw from these considerations is that to find the distribution of the coalescence time under the IIM model, we only need to find the distributions of the absorption times under the simpler processes just defined. 

A Markov process defined by the matrix $\mathbf{Q_{anc}}$, and starting in state $0$, is simply Kingman's  coalescent \citep{Kingman1982coal,Kingman1982}. For such a process, the distribution of the coalescence time is exponential, with rate equal to the inverse of the relative population size:
\begin{equation*}
f^{\left(0\right)}_{anc}(t)=\frac{1}{a}e^{-\frac{1}{a}t}, \quad \quad 0 \leq t < \infty.
\end{equation*}
A Markov process defined by $\mathbf{Q^{\left(i\right)}_{iso}}$, $i \in \left\lbrace 1,2 \right\rbrace$, is again  Kingman's coalescent, so
\begin{equation*}
f^{\left(i\right)}_{iso}(t)=\frac{1}{c_{i}}e^{-\frac{1}{c_{i}}t}, \quad \quad 0 \leq t < \infty.
\end{equation*}
Finally, with respect to the `structured' coalescent process defined by the matrix $\mathbf{Q_{mig}}$, we prove below that, for $i \in \left\lbrace1,2,3\right\rbrace$,
\begin{equation}
\label{eq:coal_time_dens_mig_stage}
f^{\left(i\right)}_{mig}(t)=-\displaystyle\sum_{j=1}^{3}V^{-1}_{ij}V_{j4} \lambda_{j}e^{-\lambda_{j}t},
\end{equation}
where $V_{ij}$ is the $(i,j)$ entry of the (non-singular) matrix $\mathbf{V}$, whose rows are the left eigenvectors of $\mathbf{Q_{mig}}$. The $(i,j)$ entry of the matrix $\mathbf{V^{-1}}$ is denoted by $V^{-1}_{ij}$. The $\lambda_{j}$ ($j \in \{1,2,3\}$) are the absolute values of those eigenvalues of $\mathbf{Q_{mig}}$ which are strictly negative (the remaining one is zero). Since the $\lambda_{j}$ are real and strictly positive, the density function of $T^{\left(i\right)}_{mig}$ is a linear combination of exponential densities.

\subsection*{Proof of (\ref{eq:coal_time_dens_mig_stage}):}

This proof has three parts. Part (i) proves the above result under two assumptions: a) $\mathbf{Q_{mig}}$ has three strictly negative eigenvalues and one zero eigenvalue, all of them real; and b) $\mathbf{Q_{mig}}$ is diagonalizable. Part (ii) proves assumption a). Part (iii) proves assumption b). To simplify the notation, we denote $\mathbf{Q_{mig}}$ by $\mathbf{Q}$ throughout the proof.

\subsubsection*{(i)}
Consider the continuous-time Markov chain defined by the matrix $\mathbf{Q}$. Let $P_{ij}(t)$, the $(i,j)$ entry of the matrix $\mathbf{P}(t)$, be the probability that the process is in state $j$ at time $t$ into the past, given that the process starts in state $i$. $\mathbf{P}(t)$ can be calculated by solving the following initial value problem:
\begin{equation*}
\begin{array}{lcl}
\mathbf{P^{'}}(t)&=&\mathbf{P}(t)\mathbf{Q}\quad;\\
\mathbf{P}(0)&=&\mathbf{I_{4}}\quad,
\end{array}
\end{equation*}
where $\mathbf{I_{4}}$ is the four by four identity matrix. Under the assumptions that $\mathbf{Q}$ is diagonalizable and that its eigenvalues are real, the solution to this initial value problem is given by: 
\begin{equation*}
\begin{array}{lcl}
\mathbf{P}(t)&=&\mathbf{P}(0)e^{\mathbf{Q}t}\\
&=&\mathbf{V^{-1}}e^{\mathbf{B}t}\mathbf{V} \quad,
\end{array}
\end{equation*}
where $\mathbf{B}$ denotes the diagonal matrix containing the real eigenvalues $\beta_{j}$, $j \in \left\lbrace 1,2,3,4 \right\rbrace$, of $\mathbf{Q}$, and $\mathbf{V}$ is the matrix of left eigenvectors of $\mathbf{Q}$. Note that $P_{i4}(t)$ is the probability that the process has reached coalescence by time $t$, if it started in state $i$. In other words, it is the \textit{cdf} of $T_{mig}^{(i)}$:
\begin{equation*}
P_{i4}(t)=F^{(i)}_{mig}(t)=\mathbf{v_{i.}^{-1}}e^{\mathbf{B}t}\mathbf{v_{.4}} \quad ,
\end{equation*}
where $\mathbf{v_{i.}^{-1}}$ is the $i^{\text{th}}$ row vector of $\mathbf{V^{-1}}$, and $\mathbf{v_{.4}}$ the $4^{\text{th}}$ column vector of $\mathbf{V}$. Differentiating, we get the \textit{pdf}:
\begin{equation*}
\begin{array}{lcl}
f^{(i)}_{mig}(t)&=&\mathbf{v_{i.}^{-1}}\mathbf{B}e^{\mathbf{B}t}\mathbf{v_{.4}}\\
&=&\displaystyle\sum_{j=1}^{4}V^{-1}_{ij}V_{j4} \beta_{j}e^{\beta_{j}t} \quad .
\end{array}
\end{equation*}
If we denote the eigenvalue equal to zero by $\beta_{4}$, and the remaining eigenvalues are strictly negative,  this \textit{pdf} can be written as a linear combination of exponential densities:
\begin{equation}
\label{eqn:pdf 2 subpopulation}
f^{(i)}_{mig}(t)=-\displaystyle\sum_{j=1}^{3}V^{-1}_{ij}V_{j4} \lambda_{j}e^{-\lambda_{j}t} \quad ,
\end{equation}
where $\lambda_{j}=|\beta_{j}|$ for $j \in \{1,2,3\}$. 

\subsubsection*{(ii)}
As $\mathbf{Q}$ is given by equation (\ref{matrix:Q2}), its characteristic polynomial, $\mathcal{P}_{\mathbf{Q}}(\beta)$, is of the form $\beta \times \mathcal{P}_{\mathbf{Q^{\left(r\right)}}}(\beta)$, where $\mathbf{Q^{\left(r\right)}}$ is the three by three upper-left submatrix of $\mathbf{Q}$, that is:
\small
\begin{equation*}
\renewcommand{\arraystretch}{1.5}
\text{\normalsize$\mathbf{Q^{\left(r\right)}}=$}
\begin{array}{c}
\begin {bmatrix} -(1+M_{1}) & M_{1}&0\\ 
M_{2}/2 &-\left(M_{1}+M_{2}\right)/2 &M_{1}/2\\
 0&M_{2} &-(1/b+M_{2})\\
\end{bmatrix}.
\end{array}
\end{equation*}
\normalsize
Thus the eigenvalues of $\mathbf{Q}$ are the solutions to  $\beta \times \mathcal{P}_{\mathbf{Q^{\left(r\right)}}}(\beta)=0$. Consequently, one of them is zero ($\beta_{4}$, say) and the remaining three eigenvalues are also eigenvalues of $\mathbf{Q^{\left(r\right)}}$.

Now consider the similarity transformation
\small
\begin{equation*}
\renewcommand{\arraystretch}{2.5}
\text{\normalsize$\mathbf{S=D Q^{(r)} D^{-1}}=$}
\begin {bmatrix} -(1+M_{1}) & \sqrt{\frac{M_{1}M_{2}}{2}}&0\\ 
\sqrt{\frac{M_{1}M_{2}}{2}} &-\left(M_{1}+M_{2}\right)/2 &\sqrt{\frac{M_{1}M_{2}}{2}}\\
 0&\sqrt{\frac{M_{1}M_{2}}{2}} &-(1/b+M_{2})\\
 \end{bmatrix},
\end{equation*} 
\begin{equation*}
\text{\normalsize where $\mathbf{D}=$}
\begin {bmatrix}\sqrt{ \frac{M_{2}}{2M_{1}}} & 0&0\\ 
0&1 &0\\
 0&0&\sqrt {\frac{M_{1}}{2M_{2}}}\\
 \end{bmatrix}\quad .
\end{equation*}
\normalsize
Because $\mathbf{S}$ is a symmetric matrix, its eigenvalues are real. Therefore, all the eigenvalues of $\mathbf{Q^{\left(r\right)}}$ are real (a similarity transformation does not change the eigenvalues). $\mathbf{S}$ is also a negative definite matrix, since its first, second and third upper-left determinants are respectively negative, positive, and negative. Hence its eigenvalues are all strictly negative, and so are the eigenvalues of $\mathbf{Q^{\left(r\right)}}$. Hence $\mathbf{Q}$ has one zero eigenvalue ($\beta_{4}$) and three real, strictly negative eigenvalues ($\beta_{1}, \beta_{2} \text{ and } \beta_{3}$).

\subsubsection*{(iii)}

Being a symmetric matrix, $\mathbf{S}$ has three independent eigenvectors. A similarity transformation preserves the number of independent eigenvectors, so $\mathbf{Q^{\left(r\right)}}$ has three independent eigenvectors as well. We denote by $\mathbf{V^{\left(r\right)}}$ the matrix whose rows are the left eigenvectors of $\mathbf{Q^{\left(r\right)}}$. 

By definition, any left eigenvector $\mathbf{v_{j.}}$ of $\mathbf{Q}$ satisfies the system of equations $\mathbf{x}(\mathbf{Q}-\mathbf{I}\beta_{j}) =\mathbf{0}$, where $\mathbf{x}=\left[x_{1}\hspace{5pt} x_{2} \hspace{5pt} x_{3} \hspace{5pt} x_{4}\right]$. The first three linear equations of this system are identical to $\mathbf{x^{\left(r\right)}}(\mathbf{Q^{\left(r\right)}}-\mathbf{I}\beta_{j}) =\mathbf{0}$, for $j \in \left\lbrace 1, 2, 3 \right\rbrace$ and $\mathbf{x^{\left(r\right)}}=\left[x_{1}\hspace{5pt} x_{2} \hspace{5pt} x_{3} \right]$, which is solved by $\mathbf{x^{\left(r\right)}}=\mathbf{v^{\left(r\right)}_{j.}}$. So this implies that, for $\beta_{j} \in \left\lbrace \beta_{1}, \beta_{2}, \beta_{3}\right\rbrace$, any row vector $\mathbf{x}$ in $\mathbb{R}^{4}$ that has $\mathbf{v^{\left(r\right)}_{j.}}$ as its first three elements will solve the first three equations of the system, whatever the value of $x_{4}$. If $x_{4}=(V^{\left(r\right)}_{j1}+\frac{1}{b}V^{\left(r\right)}_{j3})/\beta_{j}$, that vector will be an eigenvector of $\mathbf{Q}$, because it also solves the fourth equation of the system:
\small
\begin{equation*}
\left[\begin{array}{cc} \text{--------}\mathbf{v^{(r)}_{j.}}\text{--------}&\frac{V^{(r)}_{j1}+\frac{1}{b}V^{(r)}_{j3}}{\beta_{j}} \end{array} \right] 
\renewcommand\arraystretch{2} 
\renewcommand\arraycolsep{1.4pt}
\begin{array}{c}
\begin {bmatrix}
-(1+M_{1})-\beta_{j} & M_{1}&0&1\\ 
\frac{M_{2}}{2} &-\frac{\left(M_{1}+M_{2}\right)}{2}-\beta_{j} &\frac{M_{1}}{2}&0\\
 0&M_{2} &-(\frac{1}{b}+M_{2})-\beta_{j}&\frac{1}{b}\\
0&0&0&-\beta_{j}
 \end{bmatrix} 
 \end{array}
\end{equation*}
\normalsize

\begin{equation*}
=\left[\begin {array}{cccc} 0&0&0&0 \end{array} \right],
\end{equation*}
for $\beta_{j} \in \left\lbrace \beta_{1}, \beta_{2}, \beta_{3}\right\rbrace$. For the case of $\beta_{j}=\beta_{4}=0$, a row eigenvector is $\left[0\hspace{5pt} 0 \hspace{5pt} 0 \hspace{5pt} 1 \right]$. 
Collecting these row eigenvectors in a single matrix, we get $\mathbf{V}$. So,
\begin{equation*}
\renewcommand{\arraystretch}{1.5}
\mathbf{V}=
\begin {bmatrix} \text{------} &\mathbf{v^{\left(r\right)}_{1.}}&\text{------}&\frac{(V^{\left(r\right)}_{11}+\frac{1}{b}V^{\left(r\right)}_{13})}{\beta_{1}} \\ 
 \text{------} &\mathbf{v^{\left(r\right)}_{2.}} &\text{------}&\frac{(V^{\left(r\right)}_{21}+\frac{1}{b}V^{\left(r\right)}_{23})}{\beta_{2}}\\
 \text{------}&\mathbf{v^{\left(r\right)}_{3.}}& \text{------}&\frac{(V^{\left(r\right)}_{31}+\frac{1}{b}V^{\left(r\right)}_{33})}{\beta_{3}}\\
0&0&0&1
 \end{bmatrix}. 
\end{equation*}

If the matrix $\mathbf{V}$ can be shown to be invertible, then $\mathbf{Q}$ is diagonalizable. This will be the case if the system $\mathbf{xV=0}$ can only be solved by $\mathbf{x}=\left[0\hspace{5pt} 0 \hspace{5pt} 0 \hspace{5pt} 0 \right]$. Now since the three by three upper-left submatrix of $\mathbf{V}$, $\mathbf{V^{\left(r\right)}}$, is full-ranked, $x_{1}=x_{2}=x_{3}=0$ is a necessary condition for $\mathbf{xV=0}$. But then $x_{4}=0$, from the last equation of the system. Thus we have shown that $\mathbf{Q}$ is diagonalizable. $\Box$
\bigskip

We are now in position to update equations (\ref{eq:general coal dens}) and (\ref{eq:general_coal_dens_state3}) with the results just obtained. Denoting by $\mathrm{\mathbf{A}}$ the three by three matrix with entries $A_{ij}=-V^{-1}_{ij}V_{j4}$, we obtain
\begin{equation}
\label{eq:dbn_coal_time_1}
f^{\left(i\right)}_{T}(t) = \left\{
  \begin{array}{l l}
    \frac{1}{c_{i}}e^{-\frac{1}{c_{i}}t}\quad & \quad \text{for $0 \leq t \leq \tau_{1}$,}\\
   e^{-\frac{1}{c_{i}}\tau_{1}}\displaystyle\sum_{j=1}^{3} A_{ij} \lambda_{j}e^{-\lambda_{j}\left(t-\tau_{1}\right)} \quad  & \quad \text{for $\tau_{1} < t \leq \tau_{0}$,}\\
   e^{-\frac{1}{c_{i}}\tau_{1}}\displaystyle\sum_{j=1}^{3} A_{ij} e^{-\lambda_{j}\left(\tau_{0}-\tau_{1}\right)}\frac{1}{a}e^{-\frac{1}{a}\left(t-\tau_{0}\right)} & \quad \text{for $\tau_{0} < t< \infty$,}\\
0\quad &\quad \text{otherwise},
  \end{array} \right.
\end{equation}
for $i \in \left\lbrace 1,2\right\rbrace$, and
\begin{equation}
\label{eq:dbn_coal_time_2}
f^{\left(3\right)}_{T}(t) = \left\{
  \begin{array}{l l}
    \displaystyle\sum_{j=1}^{3} A_{3j} \lambda_{j}e^{-\lambda_{j}\left(t-\tau_{1}\right)} \quad  & \quad \text{for $\tau_{1} < t \leq \tau_{0}$,}\\
   \displaystyle\sum_{j=1}^{3} A_{3j} e^{-\lambda_{j}\left(\tau_{0}-\tau_{1}\right)}\frac{1}{a}e^{-\frac{1}{a}\left(t-\tau_{0}\right)} & \quad \text{for $\tau_{0} < t< \infty$,}\\
0\quad &\quad \text{otherwise}.
  \end{array} \right.
\end{equation}
If $M_{1}=M_{2}$ and $b=1$ (i.e. in the case of symmetric gene flow and equal subpopulation sizes during the gene flow period), results (\ref{eq:dbn_coal_time_1}) and  (\ref{eq:dbn_coal_time_2}) above simplify to the corresponding results in \citet{Herbots2012} -- in this case, the coefficient $A_{i3}$ in the linear combination is zero for $i \in \{1,2,3 \}$.

\subsection{Distribution of the time until coalescence under unidirectional gene flow, and in the absence of gene flow}
If either $M_{1}$ or $M_{2}$ is equal to zero, or if both are equal to zero, the above derivation of $f^{\left(i\right)}_{mig}$ is no longer applicable, as the similarity transformation in part (ii) of the proof is no longer defined  (see the denominators in some entries of the matrix $\mathbf{D}$). In this section, we derive $f^{\left(i\right)}_{mig}$, the density of the absorption time of the Markov chain defined by the matrix $\mathbf{Q_{mig}}$ given in equation (\ref{matrix:Q2}), starting from state $i$, when one or both the migration rates are zero. Again, this is all we need to fill in equations (\ref{eq:general coal dens}) and (\ref{eq:general_coal_dens_state3}) and obtain the distribution of the coalescence time of a pair of DNA sequences under the IIM model. Having gene flow in just one direction considerably simplifies the coalescent. For this reason, we resort to moment generating functions, instead of eigendecomposition, and derive fully explicit \textit{pdf}'s.

\subsubsection{Migration from subpopulation 2 to subpopulation 1 backward in time ($M_{1}=0$, $M_{2}>0$)}
\label{subsubsec:dbn_coal_time_M2}

Let $T^{\left(i\right)}_{mig}$ again be defined as the absorption time of the Markov chain generated by $\mathbf{Q_{mig}}$, now with $M_{1}=0$ and $M_{2}>0$, given that the initial state is $i \in \left\lbrace 1,2,3\right\rbrace$. A first-step argument gives the following system of equations for the \textit{mgf} of $T^{\left(i\right)}_{mig}$:
\begin{equation*}
\begin{array}{lcl}
\mathrm{E}\left[ \mathrm{exp}\left(-sT_{mig}^{\left(1\right)}\right)\right]&=&\left(\frac{1}{1+s}\right) \quad\\
\\
\mathrm{E}\left[ \mathrm{exp}\left(-sT_{mig}^{\left(2\right)}\right)\right]&=&\left(\frac{M_{2}}{1/b+M_{2}+s}\right)\mathrm{E}\left[ \mathrm{exp}\left(-sT_{mig}^{\left(3\right)}\right)\right]+\left(\frac{1/b}{1/b+M_{2}+s}\right)\quad\\
\\
\mathrm{E}\left[ \mathrm{exp}\left(-sT_{mig}^{\left(3\right)}\right)\right]&=&\left(\frac{M_{2}}{M_{2}+2s}\right)\mathrm{E}\left[ \mathrm{exp}\left(-sT_{mig}^{\left(1\right)}\right)\right] \quad.
\end{array}
\end{equation*}
Solving this system of equations and applying a partial fraction decomposition, the distributions of $T_{mig}^{(1)}$, $T_{mig}^{(2)}$ and $T_{mig}^{(3)}$ can be expressed as linear combinations of exponential distributions:
\begin{equation*}
\begin{array}{lcl}
\mathrm{E}\left[ \mathrm{exp}\left(-sT_{mig}^{\left(1\right)}\right)\right]&=&\left(\frac{1}{1+s}\right)\quad\\
\\
\mathrm{E}\left[ \mathrm{exp}\left(-sT_{mig}^{\left(2\right)}\right)\right]&=&\left(\frac{M_{2}}{1/b+M_{2}+s}\right)\left(\frac{M_{2}}{M_{2}+2s}\right)\left(\frac{1}{1+s}\right)+\left(\frac{1/b}{1/b+M_{2}+s}\right)\\
&=&\left(\frac{bM^{2}_{2}}{\left(M_{2}-2\right)\left(1-b+bM_{2}\right)}\right)\left(\frac{1}{1+s}\right)+\left(\frac{4bM_{2}}{\left(2-M_{2}\right)\left(2+bM_{2}\right)}\right)\left(\frac{M_{2}}{M_{2}+2s}\right)\\
&+&\left(\frac{1/b}{1/b+M_{2}}+\frac{b^{2}M^{2}_{2}}{\left(2+bM_{2}\right)\left(1-b+bM_{2}\right)\left(1/b+M_{2}\right)}\right)\left(\frac{1/b+M_{2}}{1/b+M_{2}+s}\right)
\quad\\
\\
\mathrm{E}\left[ \mathrm{exp}\left(-sT_{mig}^{\left(3\right)}\right)\right]&=&\left(\frac{M_{2}}{M_{2}+2s}\right)\left(\frac{1}{1+s}\right)\\
&=&\left(\frac{M_{2}}{M_{2}-2}\right)\left(\frac{1}{1+s}\right)+\left(\frac{2}{2-M_{2}}\right)\left(\frac{M_{2}}{M_{2}+2s}\right)
\quad.
\end{array}
\end{equation*}
Thus we obtain the following \textit{pdf}'s:

\begin{equation*}
\begin{array}{lcl}
f^{\left(1\right)}_{mig}\left(t\right)&=&e^{-t}\quad\\
\\
f^{\left(2\right)}_{mig}\left(t\right)&=&\left(\frac{bM^{2}_{2}}{\left(M_{2}-2\right)\left(1-b+bM_{2}\right)}\right)e^{-t}+\left(\frac{4bM_{2}}{\left(2-M_{2}\right)\left(2+bM_{2}\right)}\right)\frac{M_{2}}{2}e^{-\frac{M_{2}}{2}t}\\
&+&\left(\frac{1}{1+bM_{2}}+\frac{b^{2}M^{2}_{2}}{\left(2+bM_{2}\right)\left(1-b+bM_{2}\right)\left(1/b+M_{2}\right)}\right)\left(\frac{1}{b}+M_{2}\right)e^{-\left(1/b+M_{2}\right)t} \quad\\
\\
f^{\left(3\right)}_{mig}\left(t\right)&=&\left(\frac{M_{2}}{M_{2}-2}\right)e^{-t}+\left(\frac{2}{2-M_{2}}\right)\frac{M_{2}}{2}e^{-\frac{M_{2}}{2}t} \quad
\end{array}
\end{equation*} 
for $t>0$.

The \textit{pdf} of the coalescence time  of a pair of DNA sequences under an IIM model with $M_{1}=0$ and $M_{2}> 0$ can now be easily derived by comparing the above expressions with equation (\ref{eq:coal_time_dens_mig_stage}): $f^{(i)}_{T}(t)$ is given by equations (\ref{eq:dbn_coal_time_1}) and (\ref{eq:dbn_coal_time_2}) above, but now with
\begin{equation*}
\boldsymbol{\lambda}=\left[1 \quad \quad \frac{M_{2}}{2}\quad \quad \frac{1}{b}+M_{2}\right]\quad ,
\end{equation*}
and
\begin{equation*}
\renewcommand{\arraystretch}{1.5}
\mathbf{A}=
\begin {bmatrix} 1&0&0\\ 
\frac{bM^{2}_{2}}{\left(M_{2}-2\right)\left(1-b+bM_{2}\right)}&\frac{4bM_{2}}{\left(2-M_{2}\right)\left(2+bM_{2}\right)}&\frac{1}{1+bM_{2}}+\frac{b^{2}M^{2}_{2}}{\left(2+bM_{2}\right)\left(1-b+bM_{2}\right)\left(1/b+M_{2}\right)} \\
 \frac{M_{2}}{M_{2}-2}&\frac{2}{2-M_{2}}& 0
 \end{bmatrix}. 
\end{equation*}

\subsubsection{Migration from subpopulation 1 to subpopulation 2 backward in time ($M_{1}>0$, $M_{2}=0$)}
\label{subsubsec:dbn_coal_time_M1}
In the opposite case of unidirectional migration, and using the same derivation procedure, we find that:

\begin{equation*}
\begin{array}{lcl}
f^{\left(1\right)}_{mig}\left(t\right)&=&\left(\frac{b^{2}M_{1}^{2}}{\left(bM_{1}-2\right)\left(b-1+bM_{1}\right)}\right)\frac{1}{b}e^{-\frac{1}{b}t}+\left(\frac{4M_{1}}{\left(2-bM_{1}\right)\left(2+M_{1}\right)}\right)\frac{M_{1}}{2}e^{-\frac{M_{1}}{2}t}\\
&+&\left(\frac{1}{\left(1+M_{1}\right)}+\frac{M_{1}^{2}}{\left(2+M_{1}\right)\left(b-1+bM_{1}\right)\left(1+M_{1}\right)}\right)\left(1+M_{1}\right)e^{-\left(1+M_{1}\right)t} \quad\\
\\
f^{\left(2\right)}_{mig}\left(t\right)&=&\frac{1}{b}e^{-\frac{1}{b}t}\quad\\
\\
f^{\left(3\right)}_{mig}\left(t\right)&=&\left(\frac{bM_{1}}{bM_{1}-2}\right)\frac{1}{b}e^{-\frac{1}{b}t}+\left(\frac{2}{2-bM_{1}}\right)\frac{M_{1}}{2}e^{-\frac{M_{1}}{2}t} \quad.
\end{array}
\vspace{0.5cm}
\end{equation*} 
As a result, the \textit{pdf} of the coalescence time of a pair of sequences under the IIM model, $f^{(i)}_{T}(t)$, is again given by equations (\ref{eq:dbn_coal_time_1}) and (\ref{eq:dbn_coal_time_2}), now with
\begin{equation*}
\boldsymbol{\lambda}=\left[\frac{1}{b}\quad \quad \frac{M_{1}}{2}\quad \quad 1+M_{1} \right]
\end{equation*}
and
\begin{equation*}
\renewcommand{\arraystretch}{1.5}
\mathbf{A}=
\begin {bmatrix} 
\frac{b^{2}M^{2}_{1}}{\left(bM_{1}-2\right)\left(b-1+bM_{1}\right)}&\frac{4M_{1}}{\left(2-bM_{1}\right)\left(2+M_{1}\right)}&\frac{1}{1+M_{1}}+\frac{M^{2}_{1}}{\left(2+M_{1}\right)\left(b-1+bM_{1}\right)\left(1+M_{1}\right)} \\
1&0&0\\ 
 \frac{bM_{1}}{bM_{1}-2}&\frac{2}{2-bM_{1}}& 0
 \end{bmatrix}.
\end{equation*}

\subsubsection{Distribution of the time until coalescence under an IIM model with $M_{1}=M_{2}=0$}
\label{subsubsec:dbn_coal_time_no_gene_flow}

In this case, the IIM model reduces to a complete isolation model where both descendant populations may change size at time $\tau_{1}$ into the past. The distribution of the absorption time $T_{mig}^{(i)}$ corresponding to $\mathbf{Q_{mig}}$ will now be either exponential, if both sampled sequences are from the same subpopulation (i.e. for $i \in \{1,2\}$), or coalescence will not be possible at all until the ancestral population is reached, if we take a sequence from each subpopulation (i.e. if $i=3$). It follows that the \textit{pdf} of the coalescence time of a pair of sequences in the IIM model is given by equations (\ref{eq:dbn_coal_time_1}) and (\ref{eq:dbn_coal_time_2}) with 
\begin{equation*}
\boldsymbol{\lambda}=\left[1 \quad \quad \frac{1}{b}\quad \quad 0 \right]
\end{equation*}
and
\begin{equation*}
\renewcommand{\arraystretch}{1.5}
\renewcommand\arraycolsep{0.4cm}
\mathbf{A}=
\begin {bmatrix} 1&0&0\\ 
0&1&0 \\
 0&0&1
 \end{bmatrix}.
\end{equation*}
 
\subsection{The distribution of the number $S$ of segregating sites}

Let $S^{(i)}$ denote the number of segregating sites in a random sample of two sequences from a given locus, when the ancestral process of these sequences follows the coalescent under the IIM model and the initial state is state $i$ ($i \in \{1,2,3\}$). Recall the infinite sites assumption and assume that the distribution of the number of mutations hitting one sequence in a single generation is Poisson with mean $\mu$. As before, time is measured in units of $2N$ generations and we use the coalescent approximation. Given the coalescence time $T^{(i)}$ of two sequences, $S^{(i)}$ is Poisson distributed with mean $\theta T^{(i)}$, where $\theta=4N\mu$ denotes the scaled mutation rate. Since the \textit{pdf} of $T^{(i)}$, $f^{(i)}_{T}$, is known, the likelihood $L^{(i)}$ of an observation from a single locus corresponding to the initial state $i$ can be derived by integrating out $T^{(i)}$:

\begin{equation*}
L^{(i)}(\boldsymbol{\gamma}, \theta;s)=P\left(S^{(i)}=s;\boldsymbol{\gamma}, \theta\right)=\int\limits_0^\infty P\left(S^{(i)}=s|T^{(i)}=t\right)f^{(i)}_{T}(t)\mathrm{d}t,
\end{equation*}
where $\boldsymbol{\gamma}$ is the vector of parameters of the coalescent under the IIM model, that is, $\boldsymbol{\gamma}=\left(a, b, c_{1}, c_{2}, \tau_{1}, \tau_{0}, M_{1}, M_{2}\right)$ . There is no need to compute this integral numerically: because $f^{(i)}_{T}$ has been expressed in terms of a piecewise linear combination of exponential or shifted exponential densities, we can use standard results for a Poisson process superimposed onto an exponential or shifted exponential distribution. 

The equations (18) and (29) of \citet{Herbots2012} use this superimposition of processes to derive the distribution of $S$ under a mathematically much simpler IIM model with symmetric migration and equal subpopulation sizes during the period of migration. These equations can now be adapted to obtain the \textit{pmf} of $S$ under each of the migration scenarios dealt with in this paper. The changes accommodate the fact that the density of the coalescence time during the migration stage of the model is now given by a different linear combination of exponential densities, where the coefficients in the linear combination, as well as the parameters of the exponential densities, are no longer the same.
The \textit{pmf} of $S$ has the following general form:

\begin{equation}
\label{eq:likelihood1}
\begin{array}{lcl}
P(S^{(i)}=s)&=&\frac{\left(c_{i}\theta\right)^{s}}{\left(1+c_{i}\theta\right)^{s+1}}\left(1-e^{-\tau_{1}\left(\frac{1}{c_{i}}+\theta \right)}\displaystyle\sum_{l=0}^{s}\frac{\left(\frac{1}{c_{i}}+\theta \right)^{l}\tau_{1}^{l}}{l!}\right)\\
&&+e^{-\frac{1}{c_{i}}\tau_{1}}\displaystyle\sum_{j=1}^{3}A_{ij}\frac{\lambda_{j}\theta^{s}}{\left(\lambda_{j}+\theta\right)^{s+1}}\left(e^{-\theta\tau_{1}}\displaystyle\sum_{l=0}^{s}\frac{\left(\lambda_{j}+\theta\right)^{l}\tau_{1}^{l}}{l!}\right.\\
&&\left.-e^{-\lambda_{j}\left(\tau_{0}-\tau_{1}\right)-\theta\tau_{0}}\displaystyle\sum_{l=0}^{s}\frac{\left(\lambda_{j}+\theta\right)^{l}\tau_{0}^{l}}{l!}\right)\\
&&+\frac{e^{-\frac{1}{c_{i}}\tau_{1}-\theta\tau_{0}}\left(a\theta\right)^{s}}{\left(1+a\theta\right)^{s+1}}\left(\displaystyle\sum_{l=0}^{s}\frac{\left(\frac{1}{a}+\theta\right)^{l}\tau_{0}^{l}}{l!}\right)\displaystyle\sum_{j=1}^{3}A_{ij}e^{-\lambda_{j}\left(\tau_{0}-\tau_{1}\right)} \qquad
\end{array}
\end{equation}
for $i\in \left\lbrace1,2\right\rbrace$ and 
\begin{equation}
\label{eq:likelihood2}
\begin{array}{lcl}
P(S^{(3)}=s)&=&\displaystyle\sum_{j=1}^{3}A_{3j}\frac{\lambda_{j}\theta^{s}}{\left(\lambda_{j}+\theta\right)^{s+1}}\left(e^{-\theta\tau_{1}}\displaystyle\sum_{l=0}^{s}\frac{\left(\lambda_{j}+\theta\right)^{l}\tau_{1}^{l}}{l!}\right.\\
&&-\left.e^{-\lambda_{j}\left(\tau_{0}-\tau_{1}\right)-\theta\tau_{0}}\displaystyle\sum_{l=0}^{s}\frac{\left(\lambda_{j}+\theta\right)^{l}\tau_{0}^{l}}{l!}\right)\\
&&+\frac{e^{-\theta\tau_{0}}\left(a\theta\right)^{s}}{\left(1+a\theta\right)^{s+1}}\left(\displaystyle\sum_{l=0}^{s}\frac{\left(\frac{1}{a}+\theta\right)^{l}\tau_{0}^{l}}{l!}\right)\displaystyle\sum_{j=1}^{3}A_{3j}e^{-\lambda_{j}\left(\tau_{0}-\tau_{1}\right)} \qquad

\end{array}
\end{equation}
 for $s \in \{0,1,2,3,...\}$.
As defined in Section \ref{sec:dbn_coal_time_bi_mig}, under bidirectional migration $\boldsymbol{\lambda}=\left(\lambda_{1},\lambda_{2},\lambda_{3}\right)$ is the vector of the absolute values of the strictly negative eigenvalues of $\mathbf{Q_{mig}}$ and $A_{ij}=-V^{-1}_{ij}V_{j4}$. If migration occurs in one direction only, with $M_{1}=0$ and $M_{2}>0$, the matrix $\mathrm{\mathbf{A}}$ and the vector $\boldsymbol{\lambda}$ are those given in Section \ref{subsubsec:dbn_coal_time_M2}. In the remaining cases, when $M_{1}>0$ and $M_{2}=0$ or when there is no gene flow, $\mathrm{\mathbf{A}}$ and $\boldsymbol{\lambda}$ are given in Sections \ref{subsubsec:dbn_coal_time_M1} and \ref{subsubsec:dbn_coal_time_no_gene_flow} respectively. In the special case of $M_{1}=M_{2}$ and $b=1$, equations (\ref{eq:likelihood1}) and (\ref{eq:likelihood2}) reduce to the results of \citet{Herbots2012}.

\subsection{The likelihood of a multilocus data set}
\label{sub:full_likelihood}
Recall that, for our purposes, an observation consists of the number of nucleotide differences between a pair of DNA sequences from the same locus. To jointly estimate all the parameters of the IIM model, our method requires a large set of observations on each of the three initial states (i.e. on pairs of sequences from subpopulation 1, from subpopulation 2, and from both subpopulations). To compute the likelihood of such a data set, we use the assumption that observations are independent, so we should have no more than one observation or pair of sequences per locus and there should be free recombination between loci, i.e. loci should be sufficiently far apart.

Let each locus for the initial state $i$ be assigned a label $j_{i}\in \left\lbrace 1_{i},2_{i},3_{i},..., J_{i}\right\rbrace$, where $J_{i}$ is the total number of loci associated with initial state $i$. Denote by $\theta_{j_{i}}=4N\mu_{j_{i}}$ the scaled mutation rate at locus $j_{i}$, where $\mu_{j_{i}}$ is the mutation rate per sequence per generation at that locus. Let $\theta$ denote the average scaled mutation rate over all loci and denote by $r_{j_{i}}=\frac{\theta_{j_{i}}}{\theta}$ the relative mutation rate of locus $j_{i}$. Then $\theta_{j_{i}}=r_{j_{i}} \theta$. If the relative mutation rates are known, we can represent the likelihood of the observation at locus $j_{i}$ simply by $L(\boldsymbol{\gamma},\theta;s_{j_{i}})$. By independence, the likelihood of the data set is then given by

\begin{equation}
\label{eq:estimated likelihood}
\displaystyle 
L\left(\boldsymbol{\gamma},\theta;\mathbf{s}\right)=\prod_{i=1}^{3}\prod_{j_{i}=1}^{J_{i}} L(\boldsymbol{\gamma},\theta;s_{j_{i}}) \quad.
\end{equation}

In our likelihood method, the $r_{j_{i}}$ are treated as known constants. In practice, however, the relative mutation rates at the different loci are usually estimated using outgroup sequences \citep{Yang2002,Wang2010}.

\section{Results}
\subsection{Simulated data}
\label{sec:simulations}
We generated two batches of data sets by simulation, each batch having one hundred data sets. Each data set consists of thousands of independent observations, where each observation represents the number of nucleotide differences between two DNA sequences belonging to the same locus, when the genealogy of these sequences follows an IIM model. In batch 1 each data set has 40,000 observations: 10,000 observations for initial state 1 (two sequences drawn from subpopulation 1), 10,000 for initial state 2 (two sequences drawn from subpopulation 2), and 20,000 for initial state 3 (one sequence from each of the two subpopulations). In batch 2 each data set has 800,000 observations: 200,000 for initial state 1, 200,000 for initial state 2, and 400,000 for initial state 3. 

The data sets shown in this section were generated using the following parameter values: $a=0.75$, $\theta=2$,  $b=1.25$, $c_{1}=1.5$, $c_{2}=2$, $\tau_{0}=2$ and $\tau_{1}=1$, $M_{1}=0.5$ and $M_{2}=0.75$. Each observation in a data set refers to a different genetic locus $j$, and hence was generated using a different scaled mutation rate $\theta_{j}$ for that locus. For batch 1, we first fixed the average mutation rate over all sites to be $\theta=2$. Then, a vector of 40,000 relative size scalars $r_{j}$ was randomly generated using a Gamma(15,15) distribution. The scaled mutation rate at locus $j$ was then defined to be $\theta_{j}=r_{j}\theta$, where $r_{j}$ denotes the relative mutation rate at locus $j$, that is, the relative size of $\theta_{j}$ with respect to the average mutation rate, $\theta$. All data sets in batch 1 were generated using the same vector of relative mutation rates. The generation of the 800,000 mutation rates $\theta_{j}$ used in batch 2 was carried out following the same procedure.

When fitting the IIM model to data sets generated in this manner, the relative mutation rates $r_{j}$ are included as known constants in the log-likelihood function to be maximised.   So, as far as mutation rates are concerned, only the average over all loci is estimated (i.e. the parameter $\theta$). To increase the robustness and performance of the fitting procedure (see also \citet{Herbots2015} and the references therein), we found the maximum-likelihood estimates for a reparameterised model with parameters $\theta$, $\theta_{a}=\theta a$, $\theta_{b}=\theta b$, $\theta_{c_{1}}=\theta c_{1}$, $\theta_{c_{2}}=\theta c_{2}$, $V=\theta \left(\tau_{0}-\tau_{1} \right)$, $T_{1}=\theta \tau_{1}$ , $M_{1}$ and $M_{2}$.

\begin{figure}[tbp] 
\graphicspath{ {./} }
\centering   
\includegraphics[width=14.5cm,angle=0]{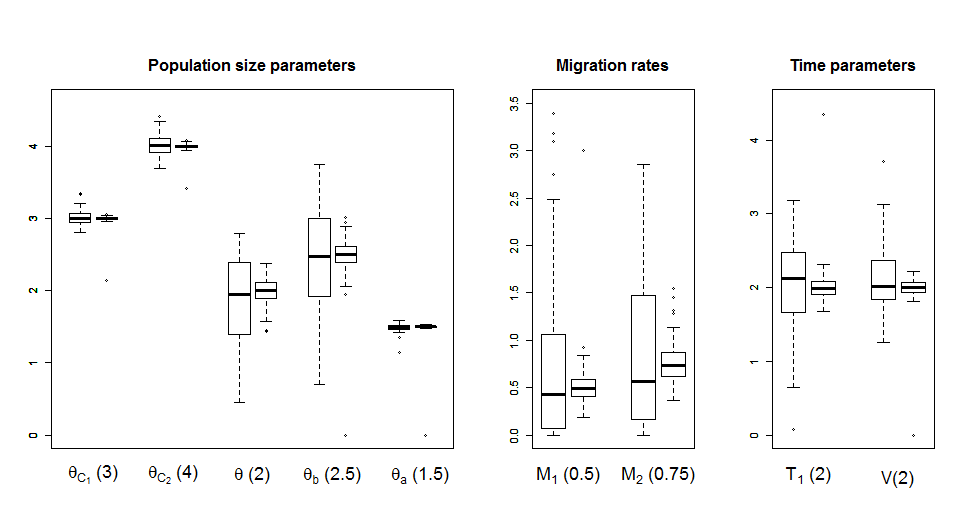} 
\caption{Estimates of population size parameters (left), migration rates (centre) and time parameters (right), for simulated data. For each parameter, the estimates shown in the left and right boxplots are based on sample sizes of \num{40000} and \num{800000} loci respectively. The values stated in parentheses are the true parameter values used to generate the data.}
\label{fig:all_bplots}
\end{figure}

The boxplots of the maximum-likelihood estimates obtained for both batches of simulated data are shown in Figure \ref{fig:all_bplots}. For each parameter, the boxplot on the left refers to batch 1 and the one on the right to batch 2. From the boxplots of time and population size parameters, it is seen that the estimates are centred around the true parameter values. Estimates for the migration rates are skewed to the right for batch 1, possibly because the true parameter values for these rates are closer to the boundary (zero) than the ones for population sizes and splitting times. For all types of parameters, increasing the sample size will decrease the variance of the maximum-likelihood estimator, as would be expected from using the correct expressions for the likelihood. In the case of the migration rate parameters, increasing the sample size eliminates most of the skewness.

\begin{figure} [tbp]
\graphicspath{ {./} }
\centering   
\includegraphics[width=14.7cm, angle=0]{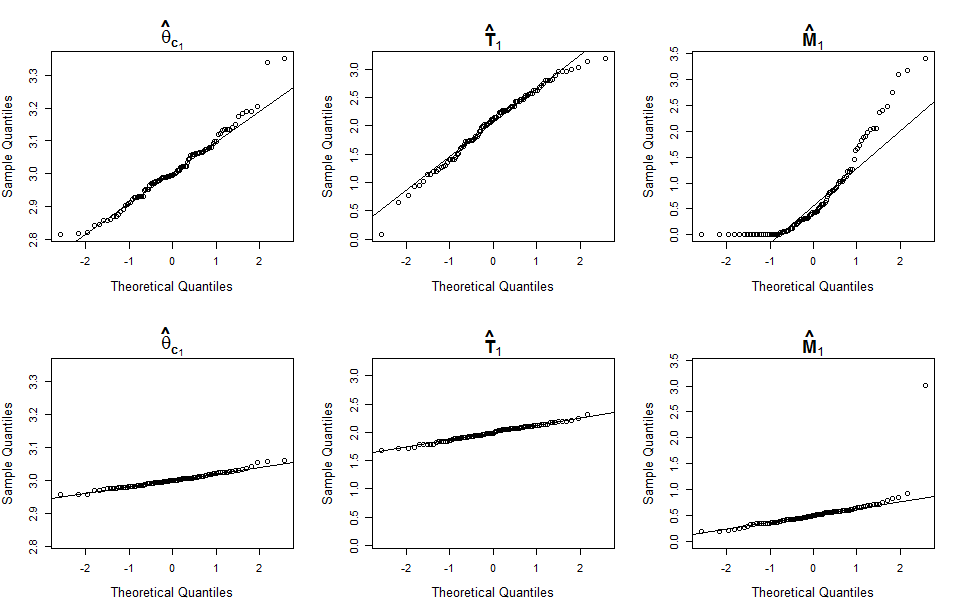} 
\caption{Normal Q-Q plots of maximum-likelihood estimates based on simulated data, for the parameters $\theta_{c_{1}}$, $T_{1}$ and $M_{1}$. The estimates shown in the top and bottom Q-Q plots are based on sample sizes of \num{40000} and \num{800000} loci respectively.}
\label{fig:normalqqplots}
\vspace{0.5cm}
\end{figure}

Figure \ref{fig:normalqqplots} shows a pair of normal Q-Q plots for each of three parameters: $\theta_{c_{1}}$ (a size parameter), $T_{1}$ (a time parameter) and $M_{1}$ (a migration parameter).  The estimates in the top and bottom Q-Q plots are based on batch 1 and batch 2 respectively.  It is clear from the figure that the distributions of the maximum-likelihood estimators of $\theta_{c_{1}}$, $T_{1}$ and $M_{1}$ become increasingly Gaussian as we increase the number of observations (compare top and bottom Q-Q plots). This is also true for the estimators of the remaining parameters (results not shown here). In addition, we can also observe that the distributions of the estimators of time and population size parameters have already a reasonably Gaussian shape for a sample size of 40,000 observations (batch 1, top plots). Again, this is true for the estimators of the remaining time and size parameters.

\subsection{The data from \citet{Wang2010}}
\label{real data}

\subsubsection{Maximum-likelihood estimation}
\label{subsubsec:ml estimation}
To illustrate our method, we apply it to a real, multilocus data set from two closely related species of \textit{Drosophila}.
The data set from \citet{Wang2010} includes an alignment of sequences covering 30247 loci, which consists of two \textit{D. simulans} assemblies, one \textit{D. melanogaster} assembly and one \textit{D. yakuba} assembly. It also includes a smaller alignment of sequences, spanning only 378 loci, which consists of two \textit{D. melanogaster} assemblies and one \textit{D. yakuba} assembly. We will call this last alignment the `Hutter subset', because it was first studied by \citet{Hutter2007}. The larger alignment will be termed the `Wang subset'. Our models are fitted to the \textit{D. melanogaster} and \textit{D. simulans} sequences from both subsets. The \textit{D. yakuba} sequences are only used as outgroup sequences, to estimate the relative mutation rates at the different loci and to calibrate time. 

To estimate the relative mutation rates $r_{j_{i}}$, we use the \textit{ad hoc} approach proposed by \citet{Yang2002}, which was also used in \citet{Wang2010} and \citet{Lohse2011}. Estimates are computed by means of the following method-of-moments estimator: 

\begin{equation}
\label{eq:method of moments}
\hat{r}_{j_{i}}=\frac{J\hspace{0.1cm}\bar{k}_{j_{i}}}{\sum^{3}_{m=1}\sum^{J_{m}}_{n=1}\bar{k}_{n_{m}}}\qquad,
\end{equation}
where $J$ is the total number of loci, and $\bar{k}_{j_{i}}$ is the average of the numbers of nucleotide differences observed in pairs of one ingroup sequence and one outgroup sequence, at locus $j_{i}$.

Since our method uses only one pair of sequences at each of a large number of independent loci,  and requires observations for all initial states, the  following procedure was adopted to select a suitable set of data. According to the genome assembly they stem from,  sequences in the Wang subset were given one of three possible tags: `Dsim1', `Dsim2' or `Dmel'. To each of the 30247 loci in the Wang subset we assigned a letter: loci with positions 1, 4, 7,... in the genome alignment were assigned the letter A; loci with positions 2, 5, 8,...  were assigned the letter B; and loci with positions 3, 6, 9,..., the letter C. A data set was then built by selecting observations corresponding to initial states 1 and 3 from the Wang subset (we used the Dsim1-Dsim2 sequences from loci A, the Dmel-Dsim1 sequences from loci B, and the Dmel-Dsim2 sequences from loci C), whilst observations corresponding to initial state 2 were obtained from the Hutter subset by comparing the two \textit{D. melanogaster} sequences available at each locus.

\begin{figure} 
\graphicspath{ {./} }
\centering      
\includegraphics[width=12.5cm]{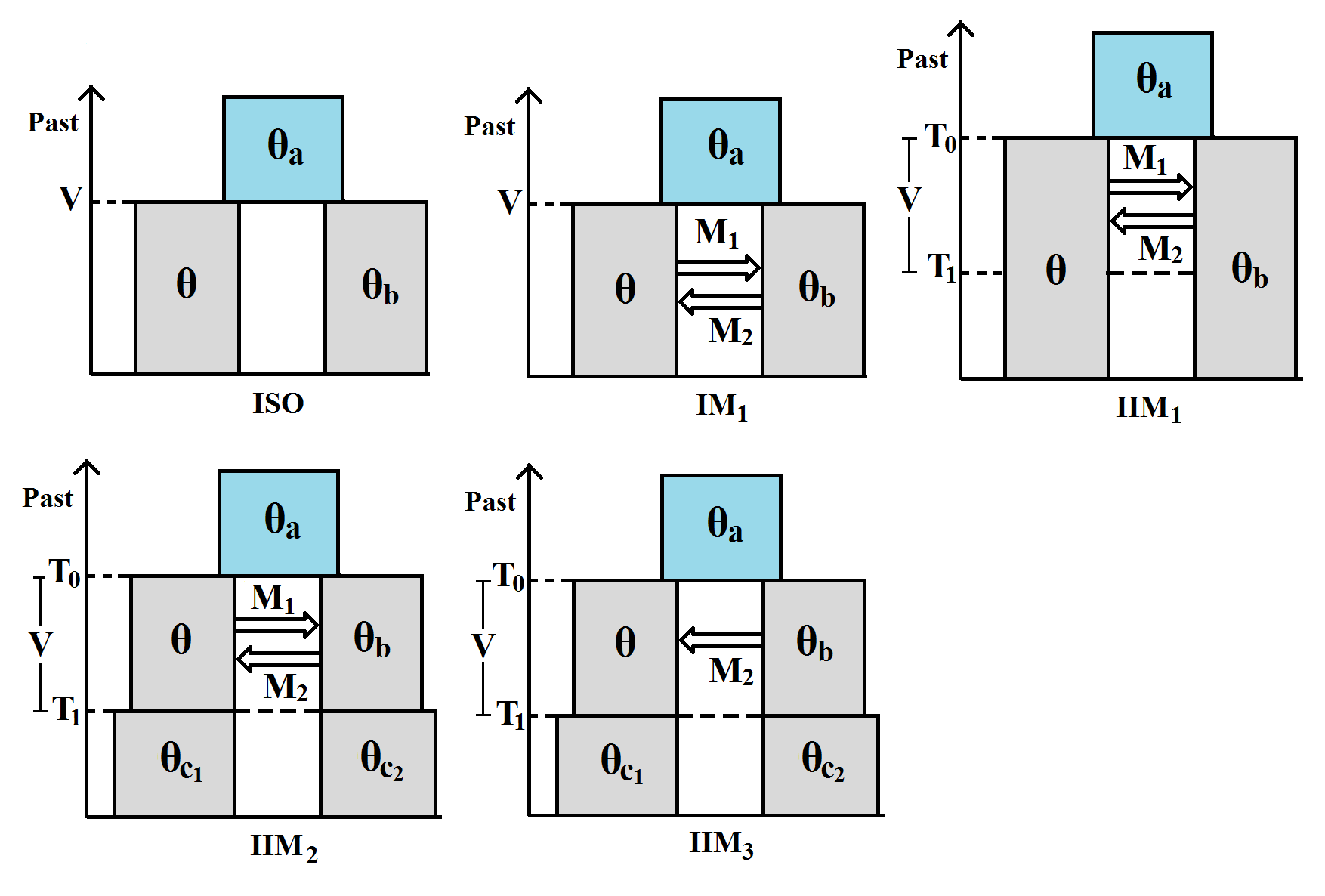} 
\caption{Models fitted to the data of \citet{Wang2010}: $\theta_{a}=\theta a$, $\theta_{b}=\theta b$, $\theta_{c_{1}}=\theta c_{1}$, $\theta_{c_{2}}=\theta c_{2}$, $V=T_{0}-T_{1}=\theta(\tau_{0}-\tau_{1})$ and $T_{1}=\theta \tau_{1}$.}
\label{fig:modeldiagrams}
\end{figure}

Table \ref{tab:ml estim} contains the  maximum-likelihood estimates for the models shown in Figure \ref{fig:modeldiagrams}. Note that the parameters of time and population size have been reparameterised as in section \ref{sec:simulations}, and recall that $M_{1}$ and $M_{2}$ are the scaled migration rates backward in time.  In the diagrams, the left and right subpopulations represent \textit{D. simulans} and \textit{D. melanogaster} respectively.  

\begin{table}
  \centering
  \begin{threeparttable}
  \scriptsize
  \caption{Results for the data of \citet{Wang2010}: maximum-likelihood estimates and values of the maximised log-likelihood, for the models shown in Figure \ref{fig:modeldiagrams}.}
    \begin{tabular}{llllllllllc}
    \toprule
    \textbf{Model} & $\boldsymbol{\theta_{a}}$ & $\boldsymbol{\theta}$ & $\boldsymbol{\theta_{b}}$ & $\boldsymbol{\theta_{c_{1}}}$ & $\boldsymbol{\theta_{c_{2}}}$ & $\mathbf{T_{1}}$ & $\mathbf{V}$ & $\mathbf{M_{1}}$ & $\mathbf{M_{2}}$ & $\mathbf{\log L(\boldsymbol{\phi})}$ \\
    \midrule
    ISO & 4.757 & 5.628 & 2.665 & -     & -     & -     & 13.705   & -     & -     & -90879.14 \\
    IM$_{1}$ & 3.974 & 5.641 & 2.493 & -     & -     & -     & 14.965  & 0.000 & 0.053 & -90276.00 \\
    IIM$_{1}$ & 3.191 & 5.581 & 2.589 & -     & -     & 6.931 & 9.928  & 0.000 & 0.528 & -90069.44 \\
    IIM$_{2}$ & 3.273 & 3.357 & 1.929 & 6.623 & 2.647 & 6.930 & 9.778  & 0.000 & 0.223 & -89899.22 \\
    IIM$_{3}$ & 3.273 & 3.357 & 1.929 & 6.623 & 2.647 & 6.930 & 9.778  & -     & 0.223 & -89899.22 \\
    \bottomrule
    \end{tabular}%
      \label{tab:ml estim}%
    \end{threeparttable}

\end{table}%

\subsubsection{Model selection}
In this section, we use likelihood ratio tests to determine which of the models listed in Table \ref{tab:ml estim} fits the data of \citet{Wang2010} best.  For reasons which we now explain, the use of such tests in the present situation is not straightforward. 

We wish to apply a standard large-sample theoretical result which states that, as the number of observations increases, the distribution of the likelihood ratio test statistic given by
\begin{equation*}
D=-2\log \lambda\left(\mathbf{s}\right) \quad,
\end{equation*}
where 
\begin{equation}
\label{eq:lrt_lambda}
\lambda\left(\mathbf{s}\right)=\frac{\displaystyle\sup_{\boldsymbol{\phi}\in \Phi_{0}}L\left(\boldsymbol{\phi}; \mathbf{s}\right)}{\displaystyle\sup_{\boldsymbol{\phi}\in \Phi}L\left(\boldsymbol{\phi}; \mathbf{s}\right)}
\end{equation}
approaches a $\chi^{2}$ distribution. In equation (\ref{eq:lrt_lambda}), $\Phi_{0}$ denotes the parameter space according to the null hypothesis ($H_{0}$). This space is a proper subspace of $\Phi$, the parameter space according to the alternative hypothesis ($H_{1}$). The number of degrees of freedom of the limiting distribution is given by the difference between the dimensions of the two spaces.

A list of sufficient regularity conditions for this result can be found, for example, in \citet[p. 516]{Casella2001}. Two of them are clearly not met in the present problem. All models under study assume independent, but \textit{not} identically distributed observations. In fact, the \textit{pmf} of an observation changes according to which species the sequences are taken from (i.e. according to the initial state) and to which locus they belong to (because each locus has its own rate of mutation). In some cases, the pairwise comparison of models is also affected by another problem: every point of $\Phi_{0}$ is a boundary point of $\Phi$. In other words, if $H_{0}$ is true, the vector of true parameters $\boldsymbol{\phi^{*}} \in \Phi_{0}$, whichever it might be, is on the boundary of $\Phi$. This irregularity is present, for example, when $M_{1}=M_{2}=0$ according to $H_{0}$   and $M_{1}$, $M_{2} \in \left[0,\infty\right) $ according to $H_{1}$.

Both these irregularities have been dealt with in the literature. As to the first one, even if observations are not identically distributed, both the maximum-likelihood estimator and the likelihood ratio test statistic still approach their respective limiting distributions if Lyapunov's condition is satisfied.  This requires the very mild assumption that $\mathrm{Var}(S_{j_{i}})$ is both bounded and bounded away from zero. It also requires \mbox{$\mathrm{E}|S_{j_{i}}-\mathrm{E}(S_{j_{i}})|^{3}$}  to be bounded \citep[pp. 233-234]{Pawitan2001}. This last condition is satisfied if $\theta_{j_{i}}$ is bounded, which again is a rather mild assumption.
  The applicability of the central limit theorem is also supported by the simulations in section \ref{sec:simulations}, which show that the distribution of the maximum-likelihood estimator indeed approaches normality as we increase the number of observations, so the lack of  identically distributed observations should not compromise the applicability of large sample results in the procedures of model selection described in this paper.

The second irregularity, that is, the problem of having  parameters on the boundary, has been the subject of papers such as \citet{Self1987} and \citet{Kopylev2010}. The limiting distribution of the likelihood ratio test statistic under this irregularity has been derived in these papers, but only for very specific cases. In most of these cases, the use of the naive $\chi_{r}^{2}$ distribution, with $r$ being the number of additional free parameters according to $H_{1}$, turns out to be conservative, because the correct null distribution is a mixture of $\chi_{\nu}^{2}$ distributions with $\nu \leq r$. Our analysis of the data of \citet{Wang2010} involves two likelihood ratio tests with parameters on the boundary (ISO vs. IM$_{1}$, and IM$_{1}$ vs. IIM$_{1}$), so we need to check that the naive $\chi_{r}^{2}$ distribution is also conservative in these cases. This was verified in a short simulation study which we now describe.

\begin{figure}[h] 
\graphicspath{ {./} }
\centering         
\includegraphics[width=14cm, angle=0]{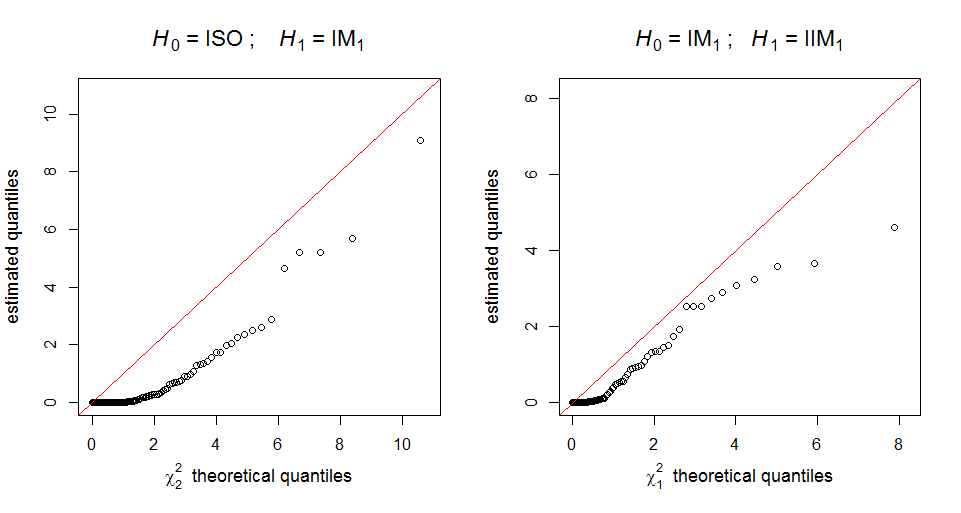} 
\caption{Q-Q plots of the estimated quantiles of the likelihood-ratio test statistic null distribution against the $\chi^{2}$ distribution theoretical quantiles. Left plot: $H_{0}$ = ISO model, $H_{1}$ = IM$_{1}$ model. Right plot: $H_{0}$ = IM$_{1}$ model, $H_{1}$ = IIM$_{1}$ model.}      
\label{fig:qq_plot_all} 
\end{figure}

We generated 100 data sets from the ISO model, each one consisting of 40,000 observations, and fitted both the ISO model ($H_{0}$) and the IM$_{1}$ model ($H_{1}$) to obtain a sample of 100 realisations of the likelihood ratio test statistic.  A Q-Q plot (Figure \ref{fig:qq_plot_all}, left boxplot) shows that the estimated quantiles of the null distribution are smaller than the corresponding theoretical quantiles of the $\chi^{2}$ distribution with 2 degrees of freedom (the difference between the dimensions of $\Phi_{0}$ and $\Phi$ in this particular case). In other words, the use of the naive $\chi^{2}$ distribution is conservative in this case.  Using $\chi^{2}_{2}$ instead of the correct null distribution, at a significance level of 5\%, the null hypothesis (i.e. the ISO model) was falsely rejected in only 1 out of the 100 simulations performed.

A similar simulation was carried out with respect to another pair of nested models: the IM$_{1}$ model (now as $H_{0}$), in which $\tau_{1}=0$, and the IIM$_{1}$ model ($H_{1}$), in which $\tau_{1} \geq 0$. Again the naive $\chi^{2}$ distribution (this time with only one degree of freedom) was found to be conservative (Figure \ref{fig:qq_plot_all}, right boxplot). And once more, only in one out of the 100 simulations performed is the null hypothesis (the IM$_{1}$ model) falsely rejected at the $5\%$ significance level, if $\chi_{1}^{2}$ is used instead of the correct null distribution.

To select the model that best fitted the data of \citet{Wang2010}, we performed the sequence of pairwise comparisons shown in Table \ref{tab:forw selec}. For any sensible significance level, this sequence of comparisons leads to the choice of IIM$_{2}$ as the best fitting model. In fact, assuming the naive $\chi^{2}$ as the null distribution, a significance level as low as $1.2 \times 10^{-74}$ is enough to reject $H_{0}$ in each of the three tests. However, since $\hat{M}_{1}=0$ for this model (see Table \ref{tab:ml estim}), a final (backward) comparison is in order: that between IIM$_{2}$ and IIM$_{3}$ (which corresponds to fixing $M_{1}$ at zero in IIM$_{2}$).  The nested model in this comparison has one parameter less and, as can be seen in Table \ref{tab:ml estim}, has the same likelihood. So, in the end, we should prefer IIM$_{3}$ to IIM$_{2}$.

\begin{table}[h]
\vspace{0.5cm}
  \centering
  \begin{threeparttable}
  \footnotesize
  \caption{Forward selection of the best model for the data of \citet{Wang2010}.}
    \begin{tabular}{llrl}
    \toprule
    $\mathbf{H_{0}}$ & $\mathbf{H_{1}}$ & $-\log 2 \lambda (\mathbf{S})$ & \textbf{P-value} \\
    \midrule
    ISO & IM$_{1}$ &603.14 & 1.147E-262 \\
    IM$_{1}$ & IIM$_{1}$ & 413.120 & 7.673E-92 \\
    IIM$_{1}$ & IIM$_{2}$ & 340.440 & 1.187E-74 \\
    \bottomrule
    \end{tabular}%
     \label{tab:forw selec}%
    \end{threeparttable}
 
\end{table}%

\subsubsection{Confidence intervals for the selected model}
The Wald confidence intervals are straightforward to calculate whenever the vector of estimates is neither on the boundary of the model's parameter space, nor too close to it. In that case, it is reasonable to assume that the vector of \textit{true} parameters does not lie on the boundary either. As a consequence, the vector of maximum-likelihood estimators is consistent and its distribution will approach a multivariate Gaussian distribution as the sample size grows \citep[see, for example,][p. 258]{Pawitan2001}. The confidence intervals can then be calculated using the inverted Hessian matrix. 

In the case of the data of \citet{Wang2010}, the vector of estimates of the selected model (IIM$_{3}$) is an interior point of the parameter space. Assuming that the vector of true parameters is also away from the boundary, we computed the Wald 95$\%$ confidence intervals shown in Table \ref{tab:best models st errors} using the inverted Hessian. In agreement with our assumption, we note that none of the confidence intervals include zero.

For large sample sizes, and for true parameter values not too close to the boundary of the parameter space, the Wald intervals are both accurate and easy to compute. To check how well the Wald intervals for the IIM$_{3}$ model fare against the more accurate \citep[see][pp. 47-48]{Pawitan2001}, but also computationally more expensive,  profile likelihood intervals, we included these in Table \ref{tab:best models st errors}.      The two methods yield very similar confidence intervals for all parameters except $\theta_{b}$. The cause of this discrepancy should lie in the fact that we only had pairs of \textit{D. melanogaster} sequences available from a few hundred loci ($\theta_{b}$ is the size of the \textit{D. melanogaster} subpopulation during the migration stage).  

\begin{table}
  \centering
  \begin{threeparttable}
  \footnotesize
  \caption{Results for the data of \citet{Wang2010}: point estimates and confidence intervals under the model IIM$_{3}$.}
    \begin{tabular}{cccc}
    \toprule
    \textbf{Parameter} & \textbf{Estimate} &\multicolumn{2}{c}{\textbf{95$\%$ Confidence intervals}} \\
             &       & \textbf{Wald} & \textbf{Profile likelihood} \\
    \midrule
    $\theta_{a}$   & 3.273 & (3.101, 3.445) & (3.100, 3.444) \\
    $\theta$ & 3.357 & (3.139, 3.575) & (3.097, 3.578) \\
    $\theta_{b}$   & 1.929 & (0.079, 3.779) & (0.672, 5.010) \\
    $\theta_{c_{1}}$ & 6.623 & (6.407, 6.839) & (6.415, 6.843) \\
    $\theta_{c_{2}}$ & 2.647 & (2.304, 2.990) & (2.331, 3.021) \\
    $T_{1}$ & 6.930 & (6.540, 7.320) & (6.542, 7.319) \\
    $V$ & 9.778 & (9.457, 10.099) & (9.456, 10.098) \\
    $M_{2}$ & 0.223 & (0.190, 0.256) & (0.186, 0.259) \\
    \bottomrule
    \end{tabular}%
    \label{tab:best models st errors}%
    \end{threeparttable}

\end{table}%

\subsubsection{Conversion of estimates}
\label{subsubsec:conv_estimates}

The conversion of the point estimates and confidence intervals to more conventional units is based on the estimates of \citet{Powell1997} of the duration of one generation ($g=0.1$ years) and the speciation time between \textit{D. yakuba} and the common ancestor of \textit{D. simulans} and \textit{D. melanogaster} (10 million years); see also \citet{Wang2010} and \citet{Lohse2011}. Using these values, we estimated $\mu$, the mutation rate per locus per generation, averaged over all loci, to be $\hat{\mu}=2.31 \times 10^{-7}$.

In Tables \ref{tab:conv pop sizes}, \ref{tab:conv div times} and \ref{tab:conv_mig_rates}, we show the converted estimates for the best fitting model IIM$_{3}$. The effective population size estimates, in units of diploid individuals, are all based on estimators of the form $\hat{N}= \frac{1}{4\hat{\mu}}\times\hat{\theta}$, for example, the estimate of the ancestral population effective size $N_{a}$ is given by $\frac{1}{4\hat{\mu}}\times\hat{\theta}_{a}$. The estimates in years of the time since the onset of speciation and of the time since the end of gene flow  are given by $\hat{t}_{0}=\frac{g}{2\hat{\mu}}\times(\hat{T}_{1}+\hat{V})$ and $\hat{t}_{1}=\frac{g}{2\hat{\mu}}\times\hat{T}_{1}$ respectively.  With respect to gene flow, we use $\hat{q}_{1}=\hat{\mu}\times\frac{\hat{M}_{2}\hat{b}}{\hat{\theta}}$ as the estimator of the \textit{fraction} of subpopulation 1 that migrates to subpopulation 2 in each generation, forward in time, and $\hat{s}_{1}=\frac{\hat{M}_{2}\hat{b}}{2}$ as the estimator of the \textit{number} of migrant sequences from subpopulation $1$ to subpopulation $2$ in each generation, also forward in time.

\begin{table}[h]
\vspace{0.5cm}
  \centering
  \begin{threeparttable}
  \footnotesize
  \caption{Effective population size estimates for the data of \citet{Wang2010} under the model IIM$_{3}$ (values in millions of diploid individuals).}
    \begin{tabular}{lccc}
    \toprule
    \textbf{Population} & \textbf{Population size} &\multicolumn{2}{c}{\textbf{95$\%$ Confidence intervals}} \\
             &       & \textbf{Wald} & \textbf{Profile likelihood} \\
    \midrule
      Ancestral population ($N_{a}$)  & 3.549  & (3.362, 3.736)& (3.362, 3.735) \\
    \textit{D. simulans}, migration stage ($N$) & 3.640  & (3.404, 3.877)& (3.359, 3.880) \\
    \textit{D. melanogaster}, migration stage ($N_{b}$) & 2.092  & (0.085, 4.099) & (0.729, 5.433)\\
    \textit{D. simulans} , isolation stage ($N_{c_{1}}$) & 7.182  & (6.949, 7.415)& (6.957, 7.421) \\
    \textit{D. melanogaster}, isolation stage ($N_{c_{2}}$) & 2.871  & (2.498, 3.243)& (2.528, 3.276) \\
    \bottomrule
    \end{tabular}%
    \label{tab:conv pop sizes}%
    \end{threeparttable}

\end{table}%

If $g$ and $\hat{\mu}$ are treated as constants, then each of the estimators just given can be expressed as a constant times a product -- or a ratio -- of the estimators of non-converted parameters. For example, we have that

\begin{equation*}
\hat{q}_{1}=\hat{\mu}\times \frac{\hat{M}_{2}\hat{b}}{\hat{\theta}}=\mathrm{constant}\times \frac{\hat{M}_{2}\hat{b}}{\hat{\theta}}\quad,
\label{eq:conv_std_error}
\end{equation*}
and
\begin{equation*}
\hat{N}_{a}= \frac{\hat{\theta}_{a}}{4\hat{\mu}}=\mathrm{constant}\times \hat{\theta}_{a}\quad.
\end{equation*}
Hence if we denote the vector of estimators of  the converted parameters by $\boldsymbol{\hat{\phi}_{c}}$, then $\boldsymbol{\hat{\phi}_{c}}=\mathbf{W}\boldsymbol{\hat{\phi}}$, where $\mathbf{W}$ is a diagonal matrix and $\boldsymbol{\hat{\phi}}=(\hat{M_{2}}\hat{b}/\hat{\theta} \, , \, \hat{\theta}_{a} \, , \,...)^{T}$. Because $\boldsymbol{\hat{\phi}}$ is a maximum-likelihood estimator (of a reparameterised model), it approaches, as the sample size increases, a multivariate Gaussian distribution with some covariance matrix $\boldsymbol{\Sigma}$; hence $\boldsymbol{\hat{\phi}_{c}}$ also approaches a multivariate Gaussian distribution, but with covariance matrix $\mathbf{W}\boldsymbol{\Sigma}\mathbf{W^{T}}$. To calculate the Wald confidence intervals of Tables \ref{tab:conv pop sizes}, \ref{tab:conv div times} and \ref{tab:conv_mig_rates}, we reparameterised the IIM model in terms of $\boldsymbol{\phi}=(M_{2}b/\theta \, , \, \theta_{a} \, , \,...)^{T}$ and found $\boldsymbol{\hat{\Sigma}}$ as the inverse of the observed Fisher information. An estimate of $\mathbf{W}\boldsymbol{\Sigma}\mathbf{W^{T}}$ followed trivially.

\begin{table}[h]
\vspace{0.5cm}
  \centering
  \begin{threeparttable}
  \footnotesize
  \caption{Divergence time estimates for the data of \citet{Wang2010} under the model IIM$_{3}$ (values in millions of years ago). }
    \begin{tabular}{lccc}
    \toprule
    \textbf{Event} & \textbf{Time since occurrence} &\multicolumn{2}{c}{\textbf{95$\%$ Confidence intervals}} \\
             &       & \textbf{Wald} & \textbf{Profile likelihood} \\
    \midrule
        Onset of speciation ($t_{0}$) & 3.624  & (3.559, 3.689)& (3.561, 3.691) \\
     Complete isolation ($t_{1}$)  & 1.503  & (1.419, 1.588)& (1.419, 1.587) \\

    \bottomrule
    \end{tabular}%
    \begin{tablenotes}
    \item Note: These are the converted estimates of $\tau_{0}$ and $\tau_{1}$ (see Figure \ref{fig:IIMfull1}).
    \end{tablenotes}
    \label{tab:conv div times}%
    \end{threeparttable}

\end{table}%

Profile likelihood confidence intervals were also computed for the parameterisation $\boldsymbol{\phi}=(M_{2}b/\theta \, , \, \theta_{a} \, ,\, ...)^{T}$. Then, if  $\mathbf{\hat{u}}$  (or $\mathbf{\hat{l}}$) is the vector of estimated upper (or lower) bounds for the parameters in $\boldsymbol{\phi}$, $\mathbf{W\hat{u}}$ (or $\mathbf{W\hat{l}}$) will be the vector of estimated upper (or lower) bounds for the converted parameters. This follows from the likelihood ratio invariance -- see, for example, \citet[p. 47-48]{Pawitan2001}.

\begin{table}[h]
  \centering
  \begin{threeparttable}
  \footnotesize
  \caption{Converted migration rates for the data of \citet{Wang2010} under the model IIM$_{3}$.}
    \begin{tabular}{lccc}
    \toprule
    \textbf{Migration parameter} & \textbf{Point Estimate} &\multicolumn{2}{c}{\textbf{95$\%$ Confidence intervals}} \\
             &       & \textbf{Wald} & \textbf{Profile likelihood} \\
    \midrule
     Migration rate ($q_{1}$)  & 8.8E-09 & (1.1E-10, 1.8E-08)& (3.2E-09, 2.4E-08) \\
     Number of migrant sequences ($s_{1}$) & 0.064  & (0.001, 0.127)& (0.023, 0.172) \\
    \bottomrule
    \end{tabular}%
    \begin{tablenotes}
    \item Note: These are forward-in-time parameters; $q_{1}$ is the fraction of  subpopulation $1$ (\textit{D. simulans}) that migrates to subpopulation $2$ (\textit{D. melanogaster}) in each generation, during the period of gene flow; $s_{1}$ is the number of sequences migrating from subpopulation $1$ to subpopulation $2$ in each generation, during the period of gene flow.
    \end{tablenotes}
    \label{tab:conv_mig_rates}%
    \end{threeparttable}

\end{table}%

\section{Discussion}

We have described a fast method to fit the isolation-with-initial-migration model to large data sets of pairwise differences at a large number of independent loci. This method relies essentially on the eigendecomposition of the generator matrix of the process during the migration stage of the model: for each set of parameter values, the computation of the likelihood involves this decomposition. Nevertheless, the whole process of estimation takes no more than a couple of minutes for a data set of tens of thousands of loci such as that of \citet{Wang2010}. The implementation of the simpler IIM model of \citet{Herbots2012}, with R code provided in \citet{Herbots2015}, is even faster than the more general method presented here, since it makes use of a fully analytical expression for the likelihood (avoiding the need for eigendecomposition of the generator matrix), but it relies on two assumptions which we have dropped here, and which are typically unrealistic for real species: the symmetry of migration rates and the equality of subpopulation sizes during the gene flow period. When compared to implementations of the IM model based on numerical integration (such as \textit{MDIV}, \textit{IM}, \textit{IMa} and \textit{IMa2}), our method is considerably faster, as it does not require the use of high-performance computing resources. It is also more appropriate for studying different species (rather than subpopulations within the same species) than any IM model method, because it drops the assumption of gene flow until the present.

Due to the number of parameters, it is not feasible to assess the performance of our method systematically over every region of the parameter space. However, our experience with simulated data sets suggests that there are two cases in which the variances of some estimators become inflated, in particular the variances of the estimators associated with the gene flow period ($\hat{M}_{1}$, $\hat{M}_{2}$, $\hat{\theta}$, $\hat{\theta}_{b}$ and $\hat{V}$). One of such cases arises whenever $V$ is very small or $T_{1}$ is very large, making it very unlikely that the genealogy of a pair of sequences under the IIM model is affected by events that occurred during the gene flow period. The second case arises when the values of the scaled migration rates are greater than one, so that the two subpopulations during the period of gene flow resemble a single panmictic population. In either of these cases, the very process of model fitting can become unstable, that is, the algorithm of maximisation of the likelihood may have difficulty converging.

It is not the goal of this paper to draw conclusions regarding the evolutionary history of \textit{Drosophila} species. We used the data of \citet{Wang2010} with the sole objective of demonstrating that our method can be applied efficiently and accurately to real data.  In Table \ref{tab:comparison WH}, we list both our estimates and those of \citet{Wang2010} for a six-parameter isolation-with-migration model (the IM$_{1}$ model -- see Figure \ref{fig:modeldiagrams}). The same table contains the estimates for our best-fitting IIM model. Our parameter estimates for the IM model agree well with those of \citet{Wang2010}. The reason that they do not match exactly lies in the fact that we have omitted the `screening procedure' described in  \citet{Wang2010} and have therefore not excluded some of the most divergent sequences in the data set. It should also be borne in mind that our model of mutation is the infinite-sites model, whereas \citet{Wang2010} have worked with the Jukes-Cantor model. Furthermore, our choice of sequence pairs was somewhat different: \citet{Wang2010} randomly selected a pair of sequences at each locus, whereas we followed the procedure described in Section \ref{subsubsec:ml estimation} above.

\begin{table}
\centering
  \begin{threeparttable}
 \caption{Comparison of converted estimates obtained with IM and IIM models}  
  \small
    \begin{tabular}{lccc}
    \toprule
            & \textbf{IM}$_{\mathrm{wh}}$ & \textbf{IM$_{1}$} & \textbf{IIM$_{3}$} \\
          \midrule
          Time since onset of speciation & 3.040 & 3.240 & 3.624 \\
    Time since isolation  & -     & -     & 1.503\vspace{2mm} \\
  
    Size of ancestral population  & 3.060 & 4.310 & 3.549 \\
       Current size of \textit{D. sim.} population & 5.990    & 6.120 & 7.182 \\
    Current size of \textit{D. mel.} population & 2.440     & 2.700     & 2.871 \\
    Size of \textit{D. sim.} population during IIM gene flow period &- & - & 3.640 \\
    Size of \textit{D. mel.} population during IIM gene flow period & - & - & 2.092\vspace{2mm}\\
    Migration rate (\textit{D. sim.} $\to$ \textit{D. mel.})   & 0.013 & 0.012 & 0.064 \\
    Migration rate (\textit{D. mel.} $\to$ \textit{D. sim.})   & 0.000 & 0.000 & - \\
    \bottomrule
    \end{tabular}%
    \begin{tablenotes}
      \scriptsize
      \item Note: Times are given in millions of years; population sizes are given in millions of individuals; the migration rates stated represent the number of sequences that migrate per generation, forward in time. The model IM$_{\mathrm{wh}}$ is the IM model fitted by \citet{Wang2010}.
    \end{tablenotes}
      \label{tab:comparison WH}%
    \end{threeparttable}
\end{table}%

There are some notable differences between the estimates for both IM models and those for the IIM model: under the IIM model, the process of speciation is estimated to have started earlier (3.6 million years ago instead of 3.0 or 3.2 million years ago), to have reached complete isolation before the present time (1.5 million years ago), and to have a higher rate of gene flow (0.064 sequences per generation instead of 0.013 or 0.012 sequences).

The method we used assumes that relative mutation rates are known (see section \ref{subsubsec:ml estimation}). In reality, we must deal with estimates of these rates, and this introduces additional uncertainty which is not reflected in the standard errors and confidence intervals obtained. However, it should be noted that, in principle, this uncertainty can be reduced to any extent desired: the method-of-moments estimator of the relative  mutation rates -- given by equation (\ref{eq:method of moments}) -- is a consistent estimator, so increasing the number of observations will get us arbitrarily close to the true value of the parameter. 
Ideally, these observations should stem from outgroup sequences only, to avoid any dependence between the estimates of relative mutation rates and the observations on ingroup pairwise differences (this was not possible here since the \citet{Wang2010} data included exactly one outgroup sequence for each locus).

Some assumptions of our IIM model, such as the infinite-sites assumption and the assumption of free recombination between loci and no recombination within loci, may not be sensible for some real data sets. The appropriateness of other assumptions, for example those regarding the constant size of populations or the constant rate of gene flow, will depend on the actual evolutionary history of the species or populations involved. How robust the IIM model is to severe violations of its assumptions is a question which lies beyond the scope of this paper. It is nevertheless obvious that, in any case, the IIM model is more robust than any model nested in it, including the IM model and models of complete isolation.

As long as the data consist of pairwise differences, the method of eigendecomposition is easily applicable to several other models of speciation. This is true for any model nested in the full IIM model; and it should also be true for any model that consists of a sequence of island models (with or without gene flow) and Wright-Fisher populations.
\vspace{\baselineskip}

\section*{Acknowledgements}

We thank Ziheng Yang for some valuable discussions and helpful suggestions. We thank Yong Wang, Jody Hey and Konrad Lohse for kindly providing the \textit{Drosophila} DNA sequence data. This research was supported by the Engineering and Physical Sciences Research Council (grant number EP/K502959/1).

\section*{Supplementary material}
In the ancillary files of this paper, we provide the R code to fit the IIM model and other simpler models, including the IM model, to data sets consisting of observations on the number of segregating sites between pairs of DNA sequences from a large number of independent loci.

\clearpage
\bibliographystyle{Chicago}
\bibliography{IIMpaper}

\end{document}